\documentstyle[twocolumn,pre,aps,epsf]{revtex}

\newcommand{\be}{\begin{equation}}
\newcommand{\ee}{\end{equation}}
\newcommand{\bea}{\begin{eqnarray}}
\newcommand{\eea}{\end{eqnarray}}
\newcommand{\Vec}[1]{\bf{#1}}

\newcommand{\bbmc}{\begin{multicols}{2} }
\newcommand{\eemc}{\end{multicols}}

\title{Critical exponents for random knots}
\author{Alexander Yu. Grosberg,${}^{1,2,3,4}$}

\address{
${}^1$Department of Physics and Center for Materials Science and
Engineering, \\ Massachusetts Institute of Technology, Cambridge,
Massachusetts 02139,  USA \\
${}^{4}${\em On leave from:\/} Institute of Biochemical Physics,
Russian Academy of Sciences, Moscow 117977, Russia
}

\address{ {\em \bigskip \begin{quote}
The size of a zero thickness (no excluded volume) polymer ring is shown to 
scale with chain length $N$ in the same way as the size of the excluded volume 
(self-avoiding) linear polymer, as $N^{\nu}$, where $\nu \approx 0.588$.  
The consequences of that fact are examined, including sizes of 
trivial and non-trivial knots.
\end{quote} }}
\begin{document}\maketitle

Knots are encountered in virtually every branch of physics, 
as evidenced by the series of books on ``Knots and Everything.'' 
Knots keep
entertaining physicists and matematicians for one and a half century, since 
W.Thomson \cite{Kelvin}.  In a more specific context, knotted polymers 
and DNA remain in the center of attention for over three decades 
\cite{delbr,frisch1,DNA}.  Nevertheless, many simple basic questions remain 
unanswered.  For instance, the most fundametal physical property of 
any macromolecule - its average size $R(N)$, scaling with chain 
length $N$, and depending on the solvent conditions - is not 
understood for knotted polymers.  In particular, the value of 
gyration radius  
determines diffusion, light scatterring, 
gel penetration and several other properties of 
macromolecules.  It is known in details for a variety of linear 
polymers under different solvent conditions.   The mere closing ends and making 
a ring-shaped chain instead of a linear one reduces its size only 
marginally, by a factor of $\sqrt{2}$ for Gaussian chain and certainly 
preserving the power law even for excluded volume one.  
The really strong dramatic effect, as we show below in this paper, 
is due to the quenched restricted set of conformations - those 
which are topologically equivalent.  Indeed, topological type 
must be quenched for all real closed ring-shaped polymers whose 
segments are unable to pass 
through each other.  The only exception is DNA in the presence of 
topo-II enzyme  \cite{topoII}, and the very fact that Nature has invented such 
marveluos tool indicates how serious is the topological problem 
for long polymers.   Thus, the formulation of the problem is very simple:  
given the polymer ring of a quenched topology, we want to estimate 
its average size.  So far, there have been only very few works on this 
problem \cite{Cloizeaux,Deutsch}, and overall the problem remains poorly understood. 
The purpose of this note is to propose some heuristic arguments addressing this 
problem.    

{\bf 1.}  In this paper, we consider the simplest ring polymer with no 
excluded volume \cite{Excl_volume}, consisting of some $N$ freely-jointed straight 
segments, length $\ell$ each.   Assume that every spatial 
shape, or conformation, of this polymer is just as likely as any other.  
In mathematical language, our ``polymer'' is simply  
a closed broken line embedded in $3D$.   With probability 1, this 
object does not have self-intersections, and, therefore, represents 
a knot.  It may be a trivial knot, if it can be continuously 
and without self-crossings transformed into a plane polygon 
topologically equivalent to a circle, or it may be one of the 
non-trivial knots. 
It has beem proven as early as 
in 1988 \cite{theorem} that the probability of a trivial knot configuration decays 
exponentially with the chain length, $N$:
\be
p_{\rm tk}(N) \simeq \exp (-N / N_0 ) \ , \label{eq:expon_probab}
\ee
where ``tk'' subscript stays for ``trivial knot'' and a constant $N_0$ 
represents the characteristic length.  This mathematical 
prediction has been beautifully confirmed in computer simulation 
\cite{Deguchi}, yielding $N_0 \simeq 340 \pm 4$.  

The restriction to broken-line polymer of straight segments is not 
necessary.  Alternatively, one can think, for instance, of a smooth worm-like 
ring object of the length $N \ell$ with effective 
segment $\ell$ \cite{worm-like}.   One can also make straight segments 
with Gaussian distributed random lengths \cite{Muthukumar}.  
The exponential result (\ref{eq:expon_probab}) holds for 
these models as well as for all similar ones, except perhaps 
with different values of $N_0$.  However, one cannot use 
continuous model (the Wiener trajectory), since it corresponds 
to the limit $N \to \infty$, $\ell \to 0$, $N \ell = {\rm const}$, 
where essentially all conformations are heavily knotted, or 
trivial knot probability is exactly zero.

While rigorous proof of the result (\ref{eq:expon_probab}) is not 
simple, it can be readily understood on a hand-waiving level.  
Indeed,  consider our polymer as a sequence of $N/g$ blobs, with 
$g$ segments in each blob.  The probability of trivial topology 
for each blob is $p_{\rm tk}(g)$, and formula (\ref{eq:expon_probab}) 
is equivalent to saying that $p_{\rm tk}(N) = 
\left[ p_{\rm tk}(g) \right] ^{N/g}$, which means that topological 
constraints between blobs are negligible compared to that between 
segments inside each blob.  The temptation is then to conclude that 
{\em vast majority of knots occur on the small length scales}.  
However appealing, this jargon statement is confusing:  indeed, 
while blob may be much smaller than the entire polymer, 
$g \ll N$, the argument above only holds if it is large 
compared 
to $N_0$ ($g \gg N_0$), which itself is numerically large.      

Exponentially small probability of a trivial knot 
(\ref{eq:expon_probab}) immeaditely explains why topological 
constraints {\it can} significantly alter the polymer size.  
Indeed, $\ell N^{1/2}$ is the size averaged over {\it all} 
conformations.  However, the size of a trivial knot is an 
average over an exponentially small {\it subset} of conformations.  
When relatively compact complex knots are removed, the remaining average 
increases.  In order to bring this idea to 
a quantitative level, a couple of further preliminary 
arguments will be handy.     

{\bf 2.} Consider now two ring polymers, with $N$ segments each. 
They form a link, trivial or non-trivial 
If rings overlap, i.e. the distance between their centers of mass 
is smaller than their sizes, the probability that the link is 
trivial vanishes, $p_{\rm tl}(N) \to 0$, 
when $N \to \infty$.  While the exact 
asymptotic expressions for $p_{\rm tl}(N)$ are not known, it 
should be qualitatively similar to the trivial knot probability   
(\ref{eq:expon_probab}), and we assume that the corresponding 
characteristic chain length $N_1$ is of the order of $N_0$. 

Speaking about the trivial link probability, one has to make 
the distinction between a link made by two rings each of 
which may be an arbitrary knot, and a link made by two trivially 
knotted rings.  Since the former probability is higher, and 
goes to 0 for large $N$, the latter, which we use below, 
goes to zero as well.  

{\bf 3.}  Since trivial link topology is highly unlikely for overlapping rings, 
the untangled ring polymers effectively present excluded volume 
for each other, even if there is no bare excluded volume for monomers. 
Simple scaling argument suggests that the excluded volume can be estimated as 
\be
v_{\rm excl}(N) \simeq R^3(N) \left[1- p_{\rm tl}(N) \right] \ ,
\label{eq:excluded_volume}
\ee 
where $R(N)$ is the (properly averaged) size of one ring.   
The concept of ``topological excluded volume'' was introduced 
in one of the first works on computer 
simulation of knots and links \cite{Maxim_link}.  It can be easily 
understood in terms of an exactly solved model 
\cite{Edwards,Frisch2} in which one polymer is a ring and the other is 
a straight line.   It has recently been rigorously proven 
\cite{Duplantier} that the excluded volume scales as 
$R^3(N)$ in $N \to \infty$ limit for two rings if their Gaussian 
linking number remains zero, even though this condition does not 
guarantee the untangled topology.  As regards 
eq. (\ref{eq:excluded_volume}), it is actually almost 
trivially correct.  Indeed, $1- p_{\rm tl}(N)$ is the probability 
for two phantom rings to be tangled, i.e., to adopt a conformation 
prohibited for real (non-phantom) untangled rings \cite{phantom}.   
 
{\bf 4.} We are now prepared to directly attack the question of 
the size, $R_{\rm tk}(N)$, of a trivial knot without excluded volume.  
Let us color two different pieces of our $N$-segment ring, both 
containing some $g \ll N$ monomers (segments).  Since neither of the 
pieces has open ends, they present uncrossable objects for one another.  
Therefore, the arguments above apply and suggest that these two 
pieces exclude for each other some volume which is of the order 
of $v_{\rm excl}(g)$.  While this identification of the pieces 
of a ring chain with separate small rings may be doubtful for 
small $g$, we expect it to become increasingly accurate with 
increasing $g$, especially when $g$ becomes greater than $N_0$.  
In that case, $p_{\rm tl}(g)$ is negligible, and we end up 
with $v_{\rm excl}(g) \simeq R^3(g)$ for $g > N_0$, which 
immediately implies that the chain of $g$-blobs belongs to the 
universality class of self-avoiding walks, and thus 
\be
R_{\rm tk}(N) \simeq R(g) \left( N/g \right)^{\nu} \ , \ \ 
\hbox{\rm where} \ \ \nu \approx 0.588 \approx 3/5 \ .
\label{eq:nu_first}
\ee 
We expect this to hold at $g>N_0$, while at smaller scales the effect 
of topological constraints is only marginal.  Thus, we 
assume $R(g) \simeq \ell g^{1/2}$ for $g$ up to about 
$N_0$ and apply the above result (\ref{eq:nu_first}) choosing 
$g \sim N_0$:
\be
R_{\rm tk}(N) \simeq 
\left\{ 
\begin{array}{rll}
\ell N^{1/2} & {\rm if} & N<N_0  \\ 
\ell N_0^{-\nu + 1/2}N^{\nu} & {\rm if} & N \gg N_0
\label{eq:trivial_knot_size}
\end{array} 
\right. \ .
\ee 
Thus, chain size in the case of trivial knot topology is controlled by 
the standard excluded volume exponent close to $3/5$.  This confirms the 
conjecture made by J. des Cloizeaux as early as in 1981 \cite{Cloizeaux}.  
We see also that this ``swollen''  
regime comes only for really long chains, because $N_0$ is 
numerically large \cite{Deguchi}. Furthermore, while in terms of 
parameters the cross-over to the ``swollen'' asymptotics is 
controlled by $N_0$, there are grounds to expect that this asymptotics 
actually develops at $N$ around a few times $N_0$ (which we expressed 
with the $\gg$ sign in formula (\ref{eq:trivial_knot_size})).   On the other 
hand, although topological constraints lead to swelling only when $N$ 
is really large, the prefactor in the second line of formula 
(\ref{eq:trivial_knot_size}) is of order unity: 
$N_0^{-\nu + 1/2} \approx 340^{-0.088} \approx 0.6$.  

It is an interesting challenge to produce a renormalization group 
analyzis for the trivial knot polymer.  One of the difficulties is 
that regular $\epsilon$-expansion techniques are hardly applicable, 
since topological constraints do not exist for linear polymers 
in dimesnions other than 3;  as a matter of fact, these 
constraints exist in the space of dimension $d$ for the objects of 
dimension $d-2$.  The manifestation of that difficulty is seen in the 
fact that trivial link probability enters the expression 
(\ref{eq:excluded_volume}) for the scale-dependent excluded volume.  
On the other hand, such RG calculation, once achieved, can shed 
a long awaited light on why $N_0$ is as large numerically as it is.     
      
{\bf 5.} Consider now some non-trivial knot, still without any excluded 
volume.   This can be addressed using the so-called ``tube inflation'' 
or ``ideal knot'' approach.  This idea has been proposed in 
several contexts by different authors, in the aspect addressing 
knot entropy it was suggested in \cite{Tube_inflation}. Recently, 
the approach received a significant 
attention \cite{Stasiak_book}.  
In particular, the present author's contribution in this book 
addresed the size of an arbitrary knot.  However, the result 
(\ref{eq:trivial_knot_size}) was not established at the time;   
instead, the size of an 
arbitrary knot was estimated under the assumption that size of a 
trivial knot scales as $N^{\mu}$ with an unknown exponent $\mu$.  We can 
now directly use the results of the author's work in 
\cite{Stasiak_book}, simply substituting $\mu = \nu$ there.  
To make the present work self-contained, we repeat the argument briefly.  

When polymer is a knotted ring, fluctuations of each segment 
are restricted by the neighboring segments.  Following the standard 
trick of the reptation theory \cite{Doi_Edwards}, we replace 
these constraints by an effective tube confining the entire 
polymer.  The difficult part of the problem is that, unlike 
the usual case of a melt \cite{Doi_Edwards},  the tube shape 
for the swollen single chain fluctuates very strongly, and 
we have to find a reasonable approximation for it.  To this end, we 
argue that there are two possibilities to achieve maximal entropy, 
corresponding to ``uniform'' and ``phase segregated'' structures.  
Let us examine both.  

We call the structure ``uniform'' when the freedom of transverse 
fluctuations is about the same everywhere 
along the polymer.  That corresponds to finding the maximal diameter, 
$D$, for which the tube of the length $L$ can be knotted into the 
given knot without self-penetration.  The shape of this ``maximally 
inflated'' tube, or the shape of its central axis line, is 
called ``ideal representation of a knot.''  It is characteristic 
of the knots topology, and so is this tube ``axis ratio'' $p=L/D$.

Thus, we replace the real topological constraints with the confinement within 
the tube whose aspect ratio, $p$, is that characteristic for 
the given knot topology.  Tube is closed, and polymer makes 
exactly one turn within the tube.  We assume further that the polymer is 
not knotted inside the tube \cite{unknot_in_tube}.  These two conditions 
together guarantee that the polymer has exactly the desirable 
topology.  We can now determine overall size of the polymer 
in space, $R$, using the following Flory-style argument:  When 
$R$ increases, polymer loses entropy because it gets stretched  
along the tube axis;  when $R$ decreases, polymer loses entropy 
because it gets squeezed perpendicular to the tube axis.  Thus, 
the equilibrium size $R$ can be conjectured to balance these two 
entropic factors.  To implement this idea, we choose $D$ and 
$L$ such that $LD^2 \sim R^3$ and, since $L/D=p$, that means 
$L \simeq R p^{2/3}$, $D \simeq R p^{-1/3}$.  We further recall 
that the unknotted polymer behaves just as an excluded 
volume one on large scales.  That allows us to employ the 
standard scaling arguments \cite{deGennes_book} to 
estimate the relevant entropy.  Specifically, transverse 
compression and longitudinal stretching in the tube are 
associated with, respectively, concentration blobs and Pincus 
tension blobs:
\be
S  \simeq  -\left( \frac{L}{N^{\nu}} 
\right)^{1/(1- \nu)} - \left( \frac{N^{3 \nu}}{L 
D^2} \right)^{1/(3 \nu -1)}  \ .
\label{eq:Interpolation}
\ee
Provided the above expressions for $L$ and $D$, this entropy has 
a maximum with respect to $R$ 
at $R \sim \ell N^{\nu} p^{-\nu + 1/3}$, 
and we identify it as the average size of the knot.  This result 
applies as long as each of the aforementioned blobs remains 
larger then $N_0$ monomers.  Equivalently, equilibrium tube 
diameter $D$ should be greater than $\ell N_0^{\nu}$, yielding 
$N>pN_0$.  For smaller blobs, Gaussian statistics should be 
valid, which means physically that we can consider a phantom 
\cite{phantom} polymer confined within the tube.  This case 
has been already considered in \cite{Tube_inflation}.  Collecting 
everything together, we obtain
\be
R_k(N) \simeq \left\{ 
\begin{array}{rll}
\ell N^{1/2} p_k^{-1/6} & {\rm if} & N < p_k N_0 \\ 
\ell N_0^{-\nu + 1/2} N^{\nu} p_k^{-\nu + 1/3} & {\rm if} & N> p_k N_0 
\end{array}
\right. \ ,
\label{eq:knot_size}
\ee
where we have included the subscript in $p_k$ to emphasize that 
this parameter is taken for the given knot $k$. 

Thus, in terms of $N$-dependence in the limit of very large $N$, 
every knot acts as an effective excluded volume, making the exponent 
equal to $\nu \approx 0.588$.  However, more complex knots, with 
larger $p$, are significantly less expanded than trivial or 
simple knots.  Furthermore, for complex knots the onset of 
swollen behavior is pushed to the range of really long chains.  These results 
are consistent with numerical observation \cite{Deguchi,Deguchi2} 
suggesting that probability of any particular knot, and not only 
trivial one, decays exponentially at large $N$ (\ref{eq:expon_probab}), 
with the characteristic length $N_0$ independent of the knot type.  
Qualitatively, one can say that increasing $N$ for the polymer 
with any given topology eventually leads to the situation 
when polymer is dominated by a very long strings in which knot 
restriction is not felt at all, and thus it should become in this 
sense equivalent to the trivial knot.  

Another way to look at the eq. (\ref{eq:knot_size}) is 
from the point of view of $p$ dependence.  For the very long 
chain, with $N \gg N_0$, topological ``memory'' acts as an 
excluded volume for relatively simple knots, with $p< N/N_0$, 
for which $R$ obeys the last line of eq. (\ref{eq:knot_size}).  
With increasing knot complexity, 
the system crosses over to the other regime, described by the 
upper line in eq. (\ref{eq:knot_size}).  Note that 
at the cross-over the chain size is about 
$R \sim \ell N^{1/3} N_0^{1/6}$.  In terms of $N$-dependence, this 
is already a collapsed chain.

Note also that the maximal value of entropy  (\ref{eq:Interpolation}) 
is about $S \sim - p$: the knot creates about one constraint 
per polymer length which is about tube diameter.  

{\bf 6.}  As we mentioned, there is an alternative way to maximize 
entropy, which we call ``phase segregated.''  To explain what we mean 
by that, consider first a prime knot.  It may be entropically favorable 
to tighten this knot as much as possible, losing virtually all the freedom 
for segments involved in the tight knot, but gaining maximal entropy 
for the rest of the chain which then fluctuates as a free unknotted 
loop.  For the composite knot, this picture 
must be slightly modified:  we still imagine a long free 
unknotted loop with several little knots - prime componets of 
the original composite knot - independently tightened in different 
places on the loop.  

To estimate roughly the entropy of such knot segregated state, 
we can use again formula (\ref{eq:Interpolation}), applying it twice 
(or several times for a composite knot): 
for the tightened part, where $p$ is roughly equal to that of the 
original knot, and for the unknotted loop part, where $p$ is 
of order one.  The chain lengths for these two are about 
$p$ and $N-p$, respectively.   Since equilibrium entropy is 
linear in $p$, this simplest estimate yields that phase 
segregated state is about as favorable entropically as the 
uniform one.  A more sophisticated approach may be needed to 
decide which of them is more stable.  It is also likely that the 
answer is sensitive to the details of local chain 
geometry (eg, freely-jointed vs worm-like segments, and the like).

{\bf 7.}  Although the qualitative statement that $p_{\rm tl}(N)$ is 
practically zero for long rings is sufficient for the main 
stream of arguments in this paper, to estimate $p_{\rm tl}(N)$ 
is itself an interesting problem.  The simple estimate 
is obtained in the following way.  Consider two strongly 
overlapping rings.
Since we are counting probabilities 
over all possible configurations of rings, regardless of their 
knottedness, the dominant statistics is Gaussian, and there 
should be about $\sqrt{N}$ contacts between the rings.  As long 
as $N$ is not very large, each contact can be thought of as  
capable to provide up to one (positive or negative) unit 
of linking.  Therefore, there are about $\sqrt{N}$ different 
linking states, with the probability of order $1/\sqrt{N}$ 
for each of them, including the untangled one.  When $N$ 
gets really very large, the situation becomes more complicated, 
because each of the $\sqrt{N}$ ``contacts'' is itself a large 
crowd of segments, which can realize a variety of linking 
arrangements.  Since entanglements in different contact regions 
do not commute to each other in this regime \cite{Nechaev_book}, 
there appears exponential diversity of different types of linkings.
Adjusting prefactors for continuity, we arrive at     
\be
p_{\rm tl}(N) \sim \left\{ 
\begin{array}{rll}
\sqrt{N_1/N} & {\rm if} & N<N_1    \\ 
\exp \left[1 - \sqrt{N/N_1} \right] & {\rm if} & N>N_1
\end{array}
\right. \ , 
\label{eq:trivial_link_probab}
\ee
where subscript ``tl'' stays for the trivial link, $N_1$ is the characteristic 
length.  The first line of this formula is in a very good agreement with 
the simulation data of the work \cite{Deguchi2} where links of the 
lengths up to 500 were examined.  This is consistent with our 
conjecture that $N_1$ is of the order of $N_0$.  It would be 
interesting to simulate longer chains, as we expect the 
$N$-dependence to cross-over to exponential in the 
range of several hundered segments.  

It is also interesting to present the blob-style argument to explain 
the lower line of the eq. (\ref{eq:trivial_link_probab}).  
Consider each ring as consisting of $N/g$ blobs, with $g$ segments 
in each blob.  About $\sqrt{N/g}$ of the blobs 
from one ring are expected to overlap with the blobs of the other 
ring.  Now the bold assumption comes (supported by an 
admittedly vague argument):  in order for two rings not to be 
linked, all of the blobs must be untangled.  Since the 
``dangerous'' blobs are only the overalpping ones, we obtain
$p_{\rm tl}(N) = \left[ p_{\rm tl}(g) \right] ^{\sqrt{N/g}}$, which 
yields the familiar answer (\ref{eq:trivial_link_probab}).

When this work has already been written, the paper \cite{Deutsch} 
appeared.  It reports simulation data for the ring-shaped 
polymer which preserves the topology of a trivial knot and does 
not have any excluded volume.  The average size of this polymer 
has been measured for a series of lengths $N=2^n$, ranging from 32 to 
2048 monomers.  The data are consistent with our results.  Specifically,
the sizes of $N$-mers were found to follow Gaussian statistics, $N^{0.5}$, for 
small $N$ ($n=5,6,7$), then power was slowly increasing with $N$, and finally  
the ratio of the sizes of 2048- and 1024-mers was close to $2^{0.58}$.  
This seems in perfect agreement with the result eq. (\ref{eq:trivial_knot_size}). 
 
The author has 
a pleasure to acknowledge useful discussions with 
B.Duplantier, I.Erukhimovich, S.Nechaev, D.Nelson, T.Witten.


\begin{thebibliography}{99}
\bibitem{Excl_volume} Excluded volume can be neglected under the condition 
$z=(d/\ell)\sqrt{N} \ll 1$ \protect\cite{book_page}, 
where $d$ is the segment diameter.  Under this condition volume exclusion 
is completely insignificant, while the very 
fact that segments cannot pass through one another is preserved, and 
so is the topology
\bibitem{worm-like}  Worm-like chain adopts smoothly curved conformations 
${\Vec r}(s)$ with the weight proportional to 
$\exp \left[ - (\ell /2) \int  \ddot{{\Vec r}} ^2 
ds \right]$, where $s$ is arc length along the chain, 
$ \ddot{{\Vec r}} = \left. d^2 {\Vec r} \right/ 
ds^2$, integration runs along the entire polymer length $N \ell$, 
and $\ell$ is called the effective Kuhn segment.
\bibitem{unknot_in_tube} The concept of an unknotted polymer confined 
within knotted tube is pretty subtle.  It can be explained in the following way.  
First, let us quench our polymer inside the tube with respect to the tube 
walls.  Then, let us pretend that the tube is phantom and bring it to the 
trivial knot non-overlapping state.  If our polymer is in a trivial knot state 
after this procedure, we say that it was unknotted witihn the tube.
\end{thebibliography}
\end{document}